\documentclass[nofootinbib,twocolumn,pre,superscriptaddress,10pt,aps,floatfix,logbibliography]{revtex4-2}
\usepackage[english]{babel}
\usepackage{graphicx}
\usepackage{amsmath}
\usepackage{amssymb}
\usepackage[colorlinks,linkcolor=blue,citecolor=blue,urlcolor=blue]{hyperref}
\usepackage{color}
\usepackage{soul}
\usepackage[normalem]{ulem}
\usepackage{xcolor}
\usepackage{makecell,booktabs}
\usepackage{multirow}
\definecolor{dblue} {RGB}{28,130,185}
\let\oldaddcontentsline\addcontentsline
\newcommand{\stoptocentries}{\renewcommand{\addcontentsline}[3]{}}
\newcommand{\starttocentries}{\let\addcontentsline\oldaddcontentsline}
\definecolor{nred}{RGB}{224,0,0}
\definecolor{nblue}{RGB}{28,130,185}
\definecolor{darkgreen}{rgb}{0,0.60,.2}
\definecolor{pgreen}{rgb}{0,1.0,0.}

\newcommand\redsout{\bgroup\markoverwith{\textcolor{brown}{\rule[0.6ex]{4pt}{1.0pt}}}\ULon}

\setcellgapes{2pt}
\makegapedcells

\newcommand{\vct}[1]{\mathbf{#1}}
\newcommand{\mrm}[1]{\mathrm{#1}}

\begin{document}
\title{Theory of two-dimensional Wigner crystals with defects:\protect\\
Interactions, melting transitions and collective modes}

\author{Pawe\l{} Matus}
\email{matus@tauex.tau.ac.il}
\affiliation{School of Physics and Astronomy, Tel Aviv University, Tel Aviv 69978, Israel}

\author{Tobias Holder}
\affiliation{School of Physics and Astronomy, Tel Aviv University, Tel Aviv 69978, Israel}

\begin{abstract}
The physics of Wigner solids is characterized by an interplay of elasticity and long-range electrostatics, endowing crystal defects with properties distinct from those in charge-neutral crystals. Recent experiments observing lattice melting and Wigner-crystal-adjacent phases in two dimensions necessitate an examination of how defects affect the long-wavelength properties of such solids. Here, we use duality techniques to construct a comprehensive framework for studying the contribution of vacancies and interstitials, dislocations, and disclinations to the effective action of two-dimensional charged crystals. This allows for a systematic investigation of the interaction energies for the different combinations of defect pairs. We further study melting transitions due to defect proliferation, assessing and justifying some of the assumptions present in the literature. In the metallic Wigner crystal phase, characterized by a finite ground state density of vacancies, we find phonons with a dispersion relation that varies in an unusual way with the vacancy density. Consequently, we discuss how thermodynamic properties of such a vacancy Fermi liquid can be probed in measurements of the melting temperature and speed of sound. 
The field theory developed here can serve as a starting point in the study of anomalous Hall crystals and charge density waves with defects.
\end{abstract}
\maketitle
\stoptocentries
\section{Introduction} 

\begin{figure}
    \includegraphics[width=0.9\linewidth]{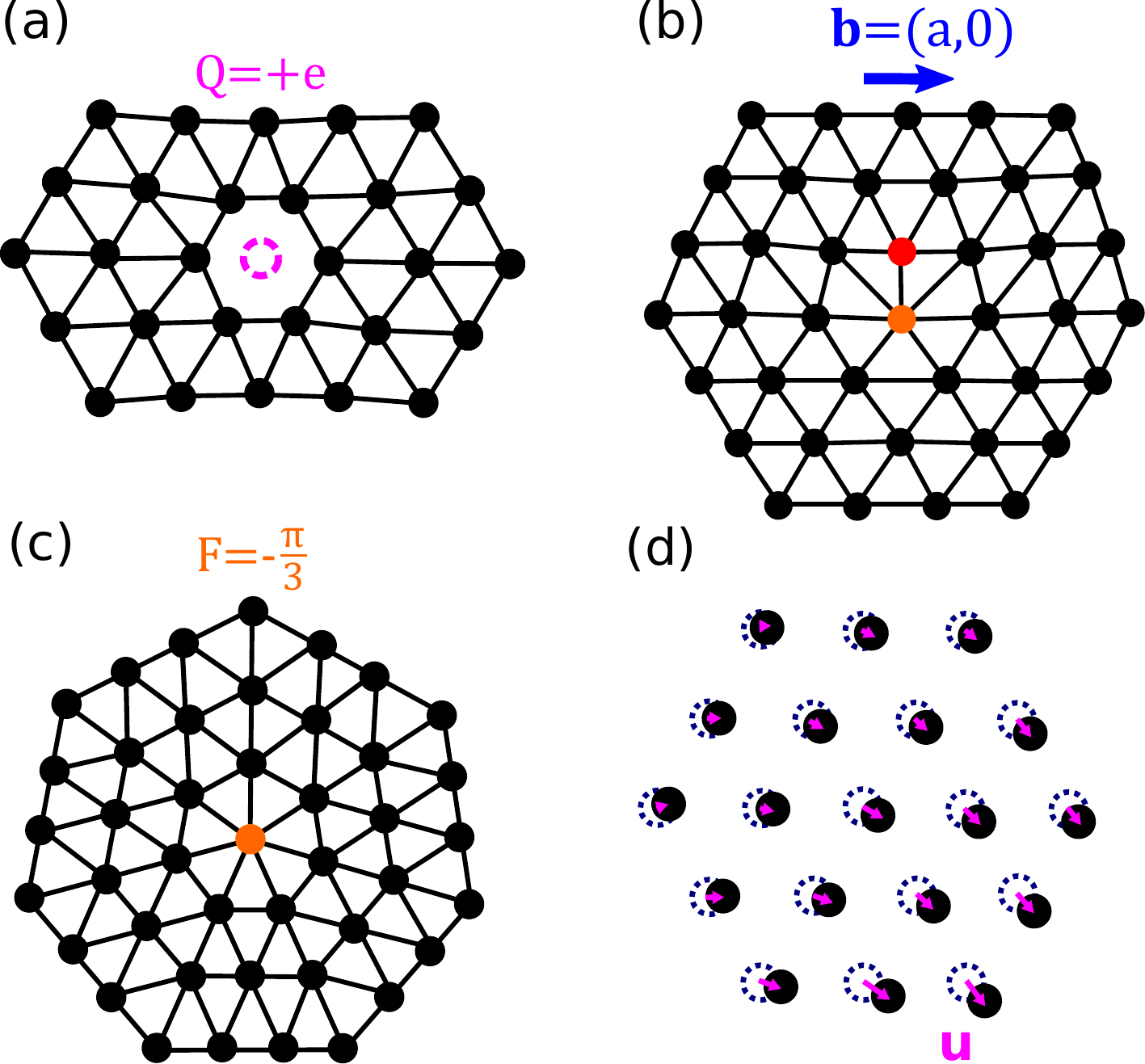}
    \caption{Crystal defects and deformations. (a) Vacancy (dashed circle) with charge $Q=e$. (b) Dislocation with Burgers vector $\vct b=(a,0)$. Lattice sites with 7 and 5 bonds are distinguished. (c) Disclination with Frank vector $F=-\frac{\pi}{3}$. Lattice site with 7 bonds is distinguished. (d) Lattice deformations are characterized by the displacement field $\vct u(t,\vct r)$, visualized with pink arrows. }
    \label{fig:deformations}
\end{figure}

Since the seminal work of Eugene Wigner in the 1930's \cite{Wigner1934,Wigner1938}, it is known that in an electron gas at very low densities Coulomb repulsion dominates over kinetic energy effects, forcing the electrons to crystallize. However, the densities required for electron crystallization proved difficult to achieve in experiment, and until 2020 it had only been observed, without the help of external magnetic field, in electron sheets on liquid helium \cite{Grimes1979} and in GaAs-based heterostructures \cite{Yoon1999}. The recent years, on the other hand, have seen convincing signatures of Wigner crystallization in an array of new two-dimensional systems such as mono- and bilayer transition metal dichalcogenides \cite{Smolenski_2021,Zhou_2021} and rhombohedral multilayer graphene \cite{Seiler2025,han2026evidencemetallicwignercrystal}, including the first direct imaging of electron crystals \cite{Xiang2025,Wang2025}. Furthermore, new related phases have been discussed, including generalized Wigner crystals \cite{Regan2020,Li2021,Padhi2021} and anomalous Hall crystals \cite{Dong_2024,Soejima2024,Zhihuan2024a,Zhihuan2024b,Patri2024,Tan_2024}. Very recently, an experiment in rhombohedral multilayer graphene has found evidence for self-doping of Wigner crystals by vacancies \cite{han2026evidencemetallicwignercrystal}, forming what is sometimes called the metallic Wigner crystal \cite{Kivelson2024,dong2026crystalscaughtdopingmetallic,feng2026selfdopedcrystalpreemptedbandinversion}. In order to interpret these observations in the context of the canonical phase diagram of a two-dimensional Wigner crystal, a more rigorous treatment of the role of lattice defects is needed. In this work, we will focus on the three types of defects pictured in Fig.~\ref{fig:deformations}: interstitials or vacancies, which can be understood as extra electrons or holes on top of the insulating crystal state, as well as topological defects known as dislocations and disclinations. 

According to the KTHNY theory \cite{Halperin1978,Strandburg1988}, thermal melting of a two-dimensional crystal is a two-step process. At the first critical temperature, the unbinding of dislocation pairs gives rise to the hexatic phase, in which a quasi-long ranged rotational order survives. Only at a higher critical temperature do the disclinations unbind as well, destroying the orientational order and producing an isotropic liquid. This scenario has been studied numerically in Wigner crystals, consistently finding that a hexatic phase can be present in a narrow window of temperatures \cite{Morf1979,Muto1999,He2003,Clark2009,Bruun2014}, but with the melting temperature deviating significantly from the KTHNY predictions and the phase transitions becoming first-order close to the quantum critical point \cite{Hwang2001,Clark2009}. There is also some experimental evidence for the hexatic phase, although not conclusive \cite{Deville1984,Kim2022}.

Vacancies and interstitials are of special importance to Wigner crystals. While a finite density of them can be found at any nonzero temperature, it has been suggested that interstitials or vacancies can proliferate even in zero temperature through a self-doping mechanism \cite{Barabanov1995,Barraza1999,Pankov2008,Kivelson2024}, a scenario that has found experimental \cite{han2026evidencemetallicwignercrystal} and theoretical \cite{dong2026crystalscaughtdopingmetallic,feng2026selfdopedcrystalpreemptedbandinversion} support in rhombohedral multilayer graphene. 

While defects in Wigner crystals have been a subject of theoretical research for a long time \cite{Fisher1979,Cockayne1990,Esfarjani1992,Zheng1994,Min-Chul_1994,Frey1994,Barabanov1995,Barraza1999,Candido2001,polini2001,Pankov2008,Irvine2012,Kivelson2022,Kivelson2024,Madathil2024,Zhuang2025}, interactions between them remain understudied. For example, stress fields produced by dislocations have typically been assumed to be the same as in charge neutral crystals with the bulk elastic modulus taken to be infinite (the incompressible limit) \cite{Fisher1979,Morf1979,Esfarjani1992}. Regarding vacancies/interstitials, previous works only contained numerics and simple power-law estimates \cite{Fisher1979,Cockayne1990,Candido2001}. Furthermore, to the best of our knowledge, interactions between different species of defects have not been studied to date.

In the present paper, we employ the technique of elasticity-gauge duality, also known as fracton-elasticity duality \cite{Kleinert1982,Kleinert1983,Zaanen_2004,Beekman_2017a,Pretko2018a,Pretko:2018jbi,Pretko2019,Caddeo_2022,Surowka2024a,Surowka2024b,Glodkowski_2025,matus2026}. As demonstrated in our recent treatment of three-dimensional charged solids in \cite{matus2026}, this technique is uniquely suited for deriving the fundamental aspects of defect physics from first principles. Within the fracton-elasticity duality framework, the different defects are viewed as sources of electromagnetic and elastic gauge fields, allowing us to find their interaction energies in a straightforward manner. We work within the paradigmatic jellium model, in which conduction electrons are treated as residing within a static uniform positively charged background, where we additionally consider possible screening of electrostatic interactions by electrodes. On the other hand, we neglect the spin. Within this setup, we corroborate some of the earlier assumptions, for example that for the purpose of studying the KTHNY melting scenario the bulk modulus can be taken as effectively infinite, or that vacancies at low densities should form a Fermi liquid.

Another advantage of the elasticity-gauge duality is that it allows for a simple modeling of defect proliferation via a Ginzburg-Landau-type approach, by which we can identify the long-wavelength modes in different phases. In particular, due to the presence of two species of charge carriers, the metallic Wigner crystal hosts a charge-neutral longitudinal phonon mode with a linear dispersion relation. The phonons of a Wigner crystal have been found to be responsible for the creation of exciton polarons \cite{zhang2025wignerpolaronsrevealwigner,wang2025spectroscopywignercrystalpolarons,adlong2025theoryexcitonpolarons2d,nyhegn2025excitoninteractingphononselectronic} and proposed to enable the detection of a pseudospin order \cite{dong2026opticaldetectionmanipulationpseudospin}.

In Sec.~\ref{sec:model}, we outline the effective action of the pure two-dimensional Wigner crystal. We then proceed to incorporate the defects using the techniques of elasticity-gauge duality.

In Sec.~\ref{sec:defects}, we find the electrostatic potential and stress fields produced by vacancies/interstitials, dislocations, and disclinations. We analyze both the case where the Coulomb interactions are screened by the presence of gates, and when they are left unscreened. These results are used to find the energies of single defects, as well as to calculate interaction energies between different defect pairs. We find that electrostatic interactions affect the results significantly.

In Sec.~\ref{sec:dislocation}, we investigate dislocation proliferation and the resulting hexatic Wigner crystal. We find that the collective modes consist of a longitudinal plasmon and a transverse phonon, similar to the pure Wigner crystal, although with a different transverse sound velocity. In the latter phase the disclination interactions are unaffected by the electronic nature of the crystal.

In Sec.~\ref{sec:vacancy}, we investigate the consequences of vacancy proliferation. We find that the proliferated holes form a Fermi liquid at low densities. As the presence of holes softens the crystal, we predict the melting temperature of the Wigner crystal to drop upon entering the metallic phase, although for realistic parameters this drop can be small, consistent with \cite{han2026evidencemetallicwignercrystal}. Furthermore, we find that the speed of sound of the longitudinal phonon depends in a nontrivial way on the defect density and their effective mass, in principle allowing to extract these parameters from the speed of sound measurements.

The following notational conventions are employed. We use Greek indices to denote both temporal and spatial tensor components, e.g., $A_\mu \in \{A_t,A_x,A_y\}$, Latin indices for spatial components only, e.g., $A_i \in \{A_x,A_y\}$, and bold symbols for spatial vectors, e.g., $\vct E = (E_x,E_y)$. We employ the Einstein summation convention for repeating indices, e.g., $A_iB_i = A_xB_x+A_yB_y$. The Fourier transform convention is $f(\omega,\vct k) = \int dtd^2x e^{i\omega t-i\vct k\cdot\vct r}f(t, \vct r)$, $f(t,\vct r) = \int \frac{d\omega d^2k}{(2\pi)^3} e^{-i\omega t+i\vct k\cdot\vct r}f(\omega, \vct k)$. Finally, the symmetric part of a tensor is denoted by round brackets: $A_{(ij)}=\frac{1}{2}\left(A_{ij}+A_{ji}\right)$, the antisymmetric part by square brackets: $A_{[ij]}=\frac{1}{2}\left(A_{ij}-A_{ji}\right)$, and the symmetric traceless part by angle brackets: $A_{\langle ij \rangle}=A_{(ij)}-\frac{1}{2}\delta_{ij}A_{kk}$.

\section{The model}
\label{sec:model}

We model the Wigner crystal as an infinitely thin, two-dimensional slab of material located at $z=0$. The displacements of the lattice sites from their equilibrium positions are described by the fields $u_i(t,\vct r)$ with $i=x,y$, see Fig.~\ref{fig:deformations}. The displacement fields $u_i(x)$ are only defined up to a translation by a lattice vector $a_i$: $u_i(x)\rightarrow u_i(x)+a_i$, allowing for the presence of topologically nontrivial defects known as dislocations. The dislocation density is defined as the curl of the gradient of $u_i$,
\begin{equation}
    \rho^{\mrm{disl}}_i = \epsilon_{jk}\partial_j\partial_k u_i,
    \label{eq:disl_density}
\end{equation}
where $i=x,y$. The surface integral of $\rho^{\mrm{disl}}_i$ defines the Burgers vector $b_i$ of the dislocation, which is a topological charge equal to a linear combination of the lattice vectors. Another topological defect is the disclination, defined as a singularity in the bond angle $\Theta = \frac{1}{2}\epsilon_{ij}\partial_iu_j$, with its density is given by
\begin{equation}
    \rho^{\mrm{dscl}}=\epsilon_{ij}\partial_i\partial_j \Theta = \frac{1}{2}\epsilon_{ij}\partial_i\rho^{\mrm{disl}}_j.
    \label{eq:dscl_density}
\end{equation}
The surface integral of $\rho^{\mrm{dscl}}$ defines the so-called Frank vector; for a $C_6$ symmetric crystal, it takes values $n\frac{2\pi}{6}$ with $n$ an integer.

In the presence of two metallic gates placed at distance $z=\pm d$ from the electron gas, the electrostatic energy of electron-electron interaction reads
\begin{equation}
    V(k) = \frac{\tanh(kd)}{2 \varepsilon k},
    \label{eq:potential}
\end{equation}
where $\varepsilon$ is the permittivity. This interaction is mediated by the electrostatic potential which we denote $A(t, \vct r)$. In addition to the lattice electrons, we allow for the presence of mobile electrons that do not belong to the lattice, which we will call vacancies and interstitials. The associated charge density is denoted $\rho^{\mathrm{vac}}(t,\vct r)$, and the corresponding coupling in the action reads $-A\rho^{\mrm{vac}}$. Taking all of the above into account, the most general quadratic action reads
\begin{equation}
\begin{split}
    S_{2D} =  &\frac12 
    \int  d t  d^2k~ \frac{2\varepsilon\, |\vct k|}{\tanh(|\vct k|d)}   A(\vct k)A(- \vct k)\\  
    + &\frac12 \int  d td^2 x
    \left[n_0m(\partial_t u_i)^2-C_{ijkl}\partial_i u_{j}\partial_ku_{l} \right.\\ &\quad\qquad\qquad\quad\left.- 2en_0u_iE_i - 2 A \rho^{\mrm{vac}}\right],
    \label{eq:action2D}
\end{split}
\end{equation}
where all fields are evaluated at the same time $t$ and the spatial dependence of fields in the second integral is left implicit. We chose to write the first part of the action in the momentum space representation since this representation offers the most clarity. In Eq.~(\ref{eq:action2D}), $m$ is the electron mass and $C_{ijkl}$ is the elasticity tensor, which for a hexagonal lattice has two components:
\begin{equation}
    C_{ijkl} = \frac{C_b}{2} \delta_{ij}\delta_{kl}+\frac{C_s}{2}(\delta_{ik}\delta_{jl}+\delta_{il}\delta_{jk}-\delta_{ij}\delta_{kl}),
    \label{eq:elastic_tensor}
\end{equation}
with $C_b$ called the bulk modulus and $C_s$ called the shear modulus. In the absence of defects, the collective modes are the transverse phonon with the dispersion
\begin{equation}
    \omega_{ph} = \sqrt{\frac{C_s}{2n_0 m}}k
    \label{eq:phonon}
\end{equation}
and the longitudinal plasmon with the dispersion
\begin{equation}
    \omega_{plasm} = \sqrt{\frac{C_s+\tilde{C}_b(k)}{2n_0 m}}k,
    \label{eq:plasmon}
\end{equation}
where the $k$-dependent bulk modulus is defined as 
\begin{equation}
    \tilde{C}_b(k) = C_b+e^2n_0^2\frac{\tanh(kd)}{\varepsilon k}.
    \label{eq:Cb_tilde}
\end{equation}
Thus, $\omega_{plasm}\propto k$ for long wavelengths $k \ll d^{-1}$, but $\omega_{plasm}\propto k^{1/2}$ when $k\gg d^{-1}$.  

To estimate the different coefficients in Eq.~(\ref{eq:action2D}), we rely on the data in  Ref.~\cite{han2026evidencemetallicwignercrystal} and take $n_0= 0.3-0.5\times10^{12}~\mrm{cm}^{-2}$, $m\approx 3m_e$, $\varepsilon = \varepsilon_r\varepsilon_0$ with $\varepsilon_r\approx 4$. Following~\cite{dong2026crystalscaughtdopingmetallic},  we assume $d=25$ nm. Regarding the magnitude of the elastic moduli, based on the fact that the only available length scale is the lattice constant $a\sim n_0^{-1/2}$ and the interaction is electrostatic, 
\begin{equation}
    C_s,C_b\sim \frac{e^2n_0^2a}{4\pi \varepsilon}.
    \label{eq:C_scaling}
\end{equation}
For example, \cite{Tozzini_1996} found $C_s \approx 0.57 \frac{e^2n_0^2a}{4\pi\varepsilon}$ and $C_b \approx 0.39 \frac{e^2n_0^2a}{4\pi\varepsilon}$ when evaluating the energy change accompanying a homogeneous deformation.

\subsection{Fracton-elasticity duality}

In the presence of dislocations and disclinations the displacement fields $u_i(t,\vct r)$ can no longer be defined via a smooth deformation of the equilibrium configuration, posing a problem for the well-posedness of the low-energy Lagrangian~(\ref{eq:action2D}). However, even in the presence of topological defects, a low-energy description can be obtained using the methods of fracton-elasticity duality \cite{Kleinert1982,Kleinert1983,Zaanen_2004,Beekman_2017a,Pretko2018a,Pretko:2018jbi,Pretko2019,Caddeo_2022,Surowka2024a,Surowka2024b,Glodkowski_2025,matus2026}. First, we apply the Hubbard-Stratonovich transformation to the elastic part of the action in Eq.~(\ref{eq:action2D}):
\begin{equation}
\begin{split}
    & S_{\mathrm{HS}} =  \frac12 
    \int  d t  d^2k~ \frac{2\varepsilon\, |\vct k|}{\tanh(|\vct k|d)}   A(\vct k)A(- \vct k)\\  
    &+
    \int dt d^2 x  \left[-\frac{1}{2n_0m}\pi_i^2+\pi_{i}\left(\partial_tu_i\right) +\frac12\tilde{C}_{ijkl}\tau_{ij}\tau_{kl} \right.\\
    &\left. +\tau_{ji}\partial_i u_{j} +\epsilon_{ij}\tau_{ij}\zeta -en_0A\partial_i u_i - A \rho^{\mrm{vac}}\right],
    \label{eq:action_HS}
\end{split}
\end{equation}
where we have introduced the Hubbard-Stratonovich fields $\pi_{i}$ and $\tau_{ij}$ interpreted as the momentum density and the stress tensor, respectively, defined the inverse elasticity tensor
\begin{equation}
    \tilde{C}_{ijkl} = \frac{C_b^{-1}}{2} \delta_{ij}\delta_{kl}+\frac{C_s^{-1}}{2}(\delta_{ik}\delta_{jl}+\delta_{il}\delta_{jk}-\delta_{ij}\delta_{kl}),
\end{equation}
and added a Lagrange multiplier $\zeta$ which constraints the stress tensor to be symmetric: $\epsilon_{ij}\tau_{ij}=0.$ The field $u_i$ is then split into the smooth part denoted $\bar{u}_i$ and the singular part $u_i^s$ related to lattice defects:
\begin{equation}
    u_i=\bar{u}_i+u_i^s.
\end{equation}
Integrating out the smooth fields $\bar{u}_i$ and $\zeta$ in the action~(\ref{eq:action_HS}) imposes the following constraints:
\begin{equation}
\begin{split}
    \partial_t\pi_i +\partial_j\tau_{ij}+en_0E_i&=0,\\
    \epsilon_{ij}\tau_{ij} &=0.
\end{split}\label{eq:constraints}
\end{equation}
All solutions to Eq.~(\ref{eq:constraints}) can be written as
\begin{equation}
\begin{split}
    \pi_i &= \epsilon_{jk}\partial_j B_{ik}, \\
    \tau_{ij} &= \epsilon_{jk}\left(\partial_k B_{it}-\partial_tB_{ik}\right)+\delta_{ij}en_0A,
\end{split} \label{eq:field_strengths}
\end{equation}
with
\begin{equation}
\begin{split}
    B_{it} &= \frac{1}{2}\epsilon_{ij}\left(\partial_j C_t-\partial_t C_{j}\right), \\
    B_{ii} &= -\frac{1}{2}\epsilon_{jk}\partial_j C_{k},
\end{split} \label{eq:field_strengths2}
\end{equation}
for some fields $B_{i\mu}$ and $C_{\mu}$. From Eqs.~(\ref{eq:field_strengths}) and~(\ref{eq:field_strengths2}) we find the gauge redundancy
\begin{equation}
\begin{split}
    B_{i\mu}&\rightarrow B_{i\mu}+\partial_\mu\lambda^{B}_i,\\
    C_\mu&\rightarrow C_\mu+\partial_\mu\lambda^{C}+2\delta_{\mu i}\epsilon_{ij}\lambda^{B}_{j},
\end{split} \label{eq:gauge}
\end{equation}
where $\lambda^{B}_i$ and $\lambda^{C}$ can be arbitrary fields. Plugging Eq.~(\ref{eq:field_strengths}) into~(\ref{eq:action_HS}) and taking into account also the singular displacement configurations $u^s_i$, we obtain the dual action
\begin{equation}
\begin{split}
    S_{\mathrm{dual}} =   
    &\frac12 
    \int  d t  d^2k~ \frac{2\varepsilon\, |\vct k|}{\tanh(|\vct k|d)}   A(\vct k)A(- \vct k)\\
    +&\int dt d^2x \left[-\frac{1}{2n_0m}\pi_i^2+\frac{1}{2}\tilde{C}_{ijkl}\tau_{ij}\tau_{kl}\right.\\
    &\left.\qquad\qquad\qquad -  A \rho^{\mathrm{vac}}+B_{i\mu}J^{\mathrm{disl}}_{i\mu}\right],
    \label{eq:action_dual}
\end{split}
\end{equation}
where $\pi_i$, $\tau_{ij}$ are given by Eqs.~(\ref{eq:field_strengths}) and~(\ref{eq:field_strengths2}), while the dislocation current $J^{\mathrm{disl}}_{i\mu}$ is given by
\begin{equation}
    J^{\mathrm{disl}}_{i\mu} = \epsilon_{\mu\nu\lambda}\partial_\nu\partial_\lambda u^s_i
\end{equation}
with the 3-dimensional Levi-Civita symbol satisfying $\epsilon_{txy}=1$.
The time component of the dislocation current is the dislocation density $J^{\mathrm{disl}}_{it}=\rho^{\mathrm{disl}}_i$ defined in Eq.~(\ref{eq:disl_density}). Furthermore, due to the identities in Eq.~(\ref{eq:field_strengths2}) we can identify the disclination coupling term $C_{\mu}J_{i\mu}^{\mathrm{dscl}}$, where
\begin{equation}
    J_t^{\mathrm{dscl}}=\frac12\epsilon_{ij}\partial_iJ^{\mathrm{disl}}_{jt}, \quad J_{i}^{\mathrm{dscl}}=-\frac12\epsilon_{ij}(\partial_tJ^{\mathrm{disl}}_{jt}+\partial_jJ^{\mathrm{disl}}_{kk}).
\end{equation}
Again, one can recognize the time component as the disclination density from Eq.~(\ref{eq:dscl_density}).

Imposing invariance of the action~(\ref{eq:action_dual}) under the different gauge transformations~(\ref{eq:gauge}) enforces the conservation of the total Burgers vector (under $\lambda^B$) and conservation of the total Frank vector of disclinations (under $\lambda^C$). Additionally, imposing electromagnetic gauge invariance enforces conservation of the total electric charge. Although this set of conservation laws will not be of primary interest to us, we note that it takes the form typical for fractonic quadrupole-conserving systems: for details, see \cite{Glodkowski_2025} and references therein.

\section{Defect fields and interactions}
\label{sec:defects}

In this section, we compute the fields produced by isolated, static defects. For reference, the equations of motion derived from the action~(\ref{eq:action_dual}) read:
\begin{subequations}
\begin{align}
    A-\frac{\tanh(kd)}{2\varepsilon\, k}\left(\rho^{\mrm{vac}}-en_0C_b^{-1}\tau_{ii}\right)&=0, \label{eq:field_vac}\\
    \partial_j\tau_{ij}-en_0\partial_iA&=0,\\\frac{1}{2}\epsilon_{ij}\epsilon_{kl}\partial_j\partial_l\left(\tilde{C}_{ikmn} \tau_{mn}\right) -\frac12\epsilon_{ij}\partial_i\rho^{\mathrm{disl}}_{j}&=0,
\end{align} \label{eq:field_eqs}
\end{subequations}
where we have set all the currents and all terms with time derivatives to zero.

\subsection{A point charge}

\begin{table}[h]
\centering
\begin{tabular}{ ccc } 
\toprule
 \textbf{Point charge $Q$}  & No gates & With gates \\
 \midrule
 Potential $A(\vct r)$  & $\dfrac{Q}{4\pi\varepsilon}l_{\mathrm{sc}}^2r^{-3}$ & $\dfrac{3Q}{4\pi\varepsilon d}\sqrt{\dfrac{\pi}{2k_*r}}e^{-k_*r}$ \\
 \midrule
 Bulk stress $\tau_{ii}(\vct r)$  & $\dfrac{C_b Q}{2\pi en_0}l_{\mathrm{sc}}r^{-3}$ & $\dfrac{3C_bQ}{2\pi e n_0 d l_{\mathrm{sc}}}\sqrt{\dfrac{\pi}{2k_*r}}e^{-k_*r}$  \\ 
 \midrule
 Shear stress $\tau_{\langle rr\rangle}(\vct r)$ & $\dfrac{C_sQ}{2\pi en_0}r^{-2}$ & $\dfrac{d}{d+l_{\mathrm{sc}}}\dfrac{C_sQ}{2\pi en_0}r^{-2}$  \\ 
 \midrule
 Shear stress $\tau_{\langle r\theta\rangle}(\vct r)$ & 0 & 0 \\
 \bottomrule
\end{tabular}
\caption{The long-range asymptotics for the different fields around a point defect with charge $Q$ in polar coordinates $(r,\theta)$.  In the right column we assume $l_{\mrm{sc}},d\ll r$, while in the middle column we set $d\rightarrow \infty$, or equivalently $l_{\mrm{sc}}\ll r \ll d$. The definitions of $l_{\mathrm{sc}}$ and $k_*$ are given in Eqs.~(\ref{eq:lsc}) and (\ref{eq:kstar}) respectively.}
\label{tab:vacancy_fields}
\end{table}

First, in order to find the fields produced by a point charge, we set $\rho^{\mrm{vac}}=Q\delta^2(\vct x)$, $\rho^{\mathrm{disl}}_{i}=0$. In the Fourier space, we find
\begin{subequations}
\begin{align}
    A(\vct k) &= \frac{Q}{2\varepsilon}\frac{l_{\rm{sc}}\tanh(kd)}{l_{\rm{sc}}k+\tanh(kd)},
    \label{eq:vacancy_potential} \\
    \tau_{ii}(\vct k) &= \frac{C_b Q}{en_0}\frac{\tanh(kd)}{l_{\rm{sc}}k+\tanh(kd)}, \label{eq:vacancy_bulk_stress}\\
    \tau_{\langle xy\rangle}(\vct k) & = \frac{C_s Q}{2en_0}\frac{2k_xk_y}{k^2}\frac{\tanh(kd)}{l_{\rm{sc}}k+\tanh(kd)},\\
    \tau_{\langle xx\rangle}(\vct k) & = \frac{C_s Q}{2en_0}\frac{k_x^2-k_y^2}{k^2}\frac{\tanh(kd)}{l_{\rm{sc}}k+\tanh(kd)}, \label{eq:vacancy_tauxx}
\end{align}
\end{subequations}
where the screening length $l_{\mathrm{sc}}^{-1}$ is
\begin{equation}
    l_{\mathrm{sc}} \equiv \frac{\varepsilon(C_s+C_b)}{e^2n_0^2}.
    \label{eq:lsc}
\end{equation}
The real-space asymptotic behavior of the different fields is summarized in Table~\ref{tab:vacancy_fields}, where we separately analyze the cases with and without the screening from the gates, the latter one obtained by replacing $\tanh(kd)\rightarrow 1$ in Eqs.~(\ref{eq:vacancy_potential})-(\ref{eq:vacancy_tauxx}). We find that when $d$ is finite, both the electrostatic potential $A$ and the bulk stress $\tau_{ii}$ are screened with the characteristic inverse length given by
\begin{equation}
    k_* \equiv \sqrt{\frac{3(d+l_{\mathrm{sc}})}{d^2 l_{\mathrm{sc}}}}.
    \label{eq:kstar}
\end{equation}
On the other hand, with no gate screening, we obtain a power-law behavior with $A,\tau_{ii}\propto r^{-3}$

More interesting is the behavior of the shear stresses, which in the polar coordinates $(x,y)=(r\cos\theta, r\sin\theta)$ can be separated into the $\tau_{\langle rr \rangle}$ and $\tau_{\langle r\theta\rangle}$ components by the transformation
\begin{equation}
    \begin{pmatrix}
        \tau_{\langle rr\rangle} \\ \tau_{\langle r\theta\rangle}
    \end{pmatrix} = 
    \begin{pmatrix}
        \cos 2\theta & \sin 2\theta \\ -\sin2\theta & \cos2\theta
    \end{pmatrix}
    \begin{pmatrix}
        \tau_{\langle xx\rangle} \\ \tau_{\langle xy\rangle}
    \end{pmatrix}.
\end{equation}
We find $\tau_{\langle r\theta\rangle}=0$ because of the radial symmetry of the problem, while $\tau_{rr}\propto r^{-2}$. This can be understood by noticing that $\tau_{ii}=-C_b\nabla\cdot\vct u$ implies that the  total charge accumulated around an interstitial or a vacancy is
\begin{equation}
    en_0\int d^2r \nabla\cdot\vct u = -en_0C_b^{-1}\tau_{ii}(\vct k=0)=-\frac{d}{d+l_{\mathrm{sc}}}Q.
\end{equation}
Therefore at large $r$ the $r$ component of the displacement vector tends to
\begin{equation}
    u_r(\vct r) \rightarrow -\frac{d}{d+l_{\mathrm{sc}}}\frac{Q}{en_0}\frac{1}{2\pi r},
\end{equation}
which in turn enforces the $r^{-2}$ scaling of the shear stress $\tau_{\langle rr\rangle}$. The energy of a single vacancy or interstitial can be separated into the energy of the core $E_{\mrm{core}}$, and the energy of this long-range displacement. We find
\begin{equation}
    E_{\mathrm{vac}} = E_{\mrm{core}} + \left(\frac{d}{d+l_{\mathrm{sc}}}\right)^2\frac{C_sQ^2}{12\pi e^2 n_0^2}a_{\mrm{core}}^{-2},
    \label{eq:Evac}
\end{equation}
where $a_{\mrm{core}}$ is the size of the core of the defect. Assuming $a_{\mrm{core}}\sim a$, $l_{\mrm{sc}}\sim a\ll d$, and taking $C_s \approx 0.57 \frac{e^2n_0^2a}{4\pi\varepsilon}$ \cite{Tozzini_1996,Bonsall1977}, we find the energy of a single vacancy/interstitial $E_{\mathrm{vac}} \approx E_{\mrm{core}} + 0.015\frac{e^2}{4\pi \varepsilon a}$.

\subsection{Dislocation}

\begin{table}[h]
\centering
\begin{tabular}{ ccc } 
\toprule
 \textbf{Dislocation $\vct b$} & No gates & With gates \\
 \midrule
 Potential $A(\vct r)$ & $\frac{-C_s b}{2 \pi e n_0}r^{-1}\sin\theta$ & $\frac{d}{d+l_{\mrm{sc}}}\frac{-C_s b}{2 \pi e n_0}r^{-1}\sin\theta$  \\
 \midrule
 Bulk stress $\tau_{ii}(\vct r)$ & $\frac{C_b C_s bl_{\mrm{sc}}}{\pi(C_b+C_s)} r^{-2}\sin\theta$ & $\alpha\frac{C_b b}{\pi}r^{-1}\sin\theta$   \\ \midrule
 Shear stress $\tau_{\langle rr\rangle}(\vct r)$ & 0 & 0 \\ \midrule
 Shear stress $\tau_{\langle r\theta\rangle}(\vct r)$  & $\frac{-C_s b}{2\pi}r^{-1}\cos\theta$ & $(1-\alpha)\frac{-C_s b}{2\pi}r^{-1}\cos\theta$ \\
 \bottomrule
\end{tabular}
\caption{The long-range asymptotics for the fields around a dislocation with Burgers vector $\vct b=b\hat{\vct x}$ in polar coordinates $(r,\theta)$. In the right column we assume $l_{\mrm{sc}},d\ll r$, while in the middle column we set $d\rightarrow \infty$, or equivalently $l_{\mrm{sc}}\ll r \ll d$. The definitions of $l_{\mathrm{sc}}$ and $\alpha$ are given in Eqs.~(\ref{eq:lsc}) and (\ref{eq:alpha}) respectively.}
\label{tab:dislocation_fields}
\end{table}

Now we consider a single dislocation with a Burgers vector $\vct b$ oriented along x, and therefore we set $\rho^{\mrm{vac}}=0$, $\rho^{\mrm{disl}}_y=0$, $\rho^{\mrm{disl}}_{x}=b\delta^2(\vct r)$. In the Fourier space, we obtain
\begin{subequations}
\begin{align}
    A(\vct k) &= i\frac{C_sb}{en_0}\frac{k_y}{k^2}\frac{\tanh(kd)}{l_{\mrm{sc}}k+\tanh(kd)}, \\
    \tau_{ii}(\vct k) & = -i \frac{2C_sC_bb}{C_s+\tilde{C}_b(k)} \frac{k_y}{k^2}, \\
    \tau_{\langle xy \rangle}(\vct k) & = i\frac{C_s\tilde{C}_b(k)b}{C_s+\tilde{C}_b(k)} \frac{2k_xk_y^2}{k^4}, \\
    \tau_{\langle xx \rangle}(\vct k) &= -i\frac{C_s\tilde{C}_b(k)b}{C_s+\tilde{C}_b(k)} \frac{k_y(k_y^2-k_x^2)}{k^4},
\end{align}
\end{subequations}
where we have used $\tilde{C}_b(k)$ defined in Eq.~(\ref{eq:Cb_tilde}). The real space asymptotic behavior of the different fields is collected in Table~\ref{tab:dislocation_fields}, in which we have denoted
\begin{equation}
    \alpha \equiv \frac{C_s}{C_s+C_b+\frac{e^2n_0^2}{\varepsilon}d}.
    \label{eq:alpha}
\end{equation}
In the presence of gates, the stresses are similar as in a neutral crystal \cite{Landau1986}, but with a renormalized bulk modulus: $C_b\rightarrow C_b +\frac{e^2n^2}{\varepsilon}d$, see Eq.~(\ref{eq:alpha}). Without gate screening, on the other hand, the bulk stress is strongly suppressed as it follows the $r^{-2}$ power law instead of $r^{-1}$. In either case, the energy of a single defect is divergent: in a system of size $L$, it grows as $E_{\mrm{disl}}\propto\ln L$.

\subsection{Disclination}

\begin{table}[h]
\centering
\begin{tabular}{ ccc } 
\toprule
 \textbf{Disclination $F$} & No gates & With gates \\
 \midrule
 Potential $A(\vct r)$ & $\frac{-C_s F}{\pi e n_0}\ln r$ & $\frac{d}{d+l_{\mrm{sc}}}\frac{-C_s F}{\pi e n_0}\ln r$ \\
 \midrule
 Bulk stress $\tau_{ii}(\vct r)$ & $\frac{-2C_b C_s F l_{\mrm{sc}}}{\pi(C_b+C_s)} r^{-1}$ & $\alpha\frac{2C_bF}{\pi}\ln r$ \\ \midrule
 Shear stress $\tau_{\langle rr\rangle}(\vct r)$ & $\frac{-C_sF}{2\pi}$ & $(1-\alpha)\frac{-C_sF}{2\pi}$ \\ \midrule
 Shear stress $\tau_{\langle r\theta\rangle}(\vct r)$ & 0 & 0 \\
 \bottomrule
\end{tabular}
\caption{The long-range asymptotics for the fields around a disclination with Frank vector $F$ in polar coordinates $(r,\theta)$. In the middle column we assume $l_{\mrm{sc}},d\ll r$, while in the right column we set $d\rightarrow \infty$, or equivalently $l_{\mrm{sc}}\ll r \ll d$. The definition of $l_{\mathrm{sc}}$ is given in Eqs.~(\ref{eq:lsc}).}
\label{tab:disclination_fields}
\end{table}

Now we consider a single disclination with the Frank vector $F$ oriented along x, and therefore we set $\rho^{\mrm{vac}}=0$, $\frac{1}{2}\epsilon_{ij}\partial_i\rho^{\mrm{disl}}_j=F\delta^{(2)}(\vct r)$. In the Fourier space, we obtain
\begin{subequations}
\begin{align}
    A(\vct k) &= \frac{2C_sF}{en_0}\frac{1}{k^2}\frac{\tanh(kd)}{l_{\mrm{sc}}k+\tanh(kd)}, \\
    \tau_{ii}(\vct k) & = -\frac{4C_sC_b F}{C_s+\tilde{C}_b(k)} \frac{1}{k^2}, \\
    \tau_{\langle xy \rangle}(\vct k) & = \frac{2C_s\tilde{C}_b(k) F}{C_s+\tilde{C}_b(k)} \frac{2k_xk_y}{k^4}, \\
    \tau_{\langle xx \rangle}(\vct k) &= \frac{2C_s\tilde{C}_b(k) F}{C_s+\tilde{C}_b(k)} \frac{k_x^2-k_y^2}{k^4}.
\end{align}
\end{subequations}
The real space asymptotic behavior of the different fields is collected in Table~\ref{tab:disclination_fields}. As for dislocations, the dependence of the stresses on $r$ is similar as in a neutral crystal \cite{deWit1973IV} with the exception that the bulk stress is strongly suppressed in the absence of gates, and the energy of the defect grows as $E_{\mrm{dscl}}\propto L^2$.

\subsection{Interaction energies}

\begin{table}[h]
\centering
\begin{tabular}{ ccc } 
 \toprule
 Defect charges & No gates & With gates \\ \midrule
 $Q_1$-$Q_2$ & $\frac{Q_1Q_2}{4\pi\epsilon}l_{\mathrm{sc}}^2r^{-3}$ & $\frac{3Q_1Q_2}{4\pi\varepsilon d}\sqrt{\frac{\pi}{2k_*r}}\exp(-k_*r)$ \\ \midrule
 $\vct b$-$Q$ & $-\frac{C_s}{2\pi e n_0}\frac{Q (\vct b\times\vct r)}{r^2}$ & $-\frac{d}{d+l_{\mrm{sc}}}\frac{C_s}{2\pi e n_0}\frac{Q (\vct b\times\vct r)}{r^2}$  \\ \midrule
 $\vct b_1$-$\vct b_2$ & $-\frac{C_s}{2\pi}(\vct b_1\cdot\vct b_2)\ln r$ & $-\frac{C_s(1-\alpha)}{2\pi}(\vct b_1\cdot\vct b_2)\ln r$ \\
 \midrule
 $F$-$Q$ & $-\frac{C_s F Q}{\pi e n_0}\ln r$ & $-\frac{d}{d+l_{\mrm{sc}}}\frac{C_s F Q}{\pi e n_0} \ln r$ \\ \midrule
 $F$-$\vct b$ & $\frac{C_s}{2\pi}F(\vct b\times\vct r) \ln r$ & $\frac{C_s(1-\alpha)}{2\pi}F(\vct b\times\vct r) \ln r$  \\ \midrule
 $F_1$-$F_2$ & $\frac{C_sF_1F_2}{\pi}r^2\ln r$ & $\frac{C_sF_1F_2(1-\alpha)}{\pi}r^2\ln r$ \\
 \bottomrule
\end{tabular}
\caption{Leading-order long range interaction energy between the different types of defects. $Q$ represents a vacancy or an interstitial, $\vct b$ a dislocation, and $F$ a disclination. $\vct r$ is the vector linking the first to the second defect in the pair.}
\label{tab:interactions}
\end{table}

Having found the fields produced by the defects, we are in the position to calculate the interaction energy. Since the problem is static, we can fix the gauge such that $B_{ik}=0$ in Eq.~(\ref{eq:field_strengths}), and solve fo the elastic potential $B_{it}$. For a vacancy, we find
\begin{equation}
    B_{it} = \frac{d}{d+l_\mrm{sc}}\frac{C_sQ}{2\pi en_0}\frac{\epsilon_{ij}x_j}{r^2}.
\end{equation}
For a dislocation, we find
\begin{equation}
    B_{it} = \frac{C_s(1-\alpha)}{2\pi}\left(b_i\ln r-\frac{(\vct b\cdot x)x_i }{r^2}\right).
\end{equation}
For a disclination, we find 
\begin{equation}
\begin{split}
    C_t &= -(1-\alpha)\frac{C_sF}{\pi}r^2\ln r,\\ 
    B_{it} & =-(1-\alpha)\frac{C_sF}{2\pi}\epsilon_{ij} x_j\ln r.
\end{split}
\end{equation}
The resulting interaction energies are collected in Table~\ref{tab:interactions}.

\section{Dislocation proliferation: hexatic Wigner crystal}
\label{sec:dislocation}

According to the KTHNY theory, a two dimensional crystal loses shear rigidity via a proliferation of dislocations at the critical temperature $T_c$,
\begin{equation}
    \frac{C_s\tilde{C}_b(k\rightarrow 0)}{C_s+\tilde{C}_b(k\rightarrow 0)}(k_b T_c)^{-1}=\frac{C_s(1-\alpha)}{k_b T_c}=\frac{4\pi}{a^2},
    \label{eq:kbThexatic}
\end{equation}
where $\tilde{C}_b(k)$ was defined in Eq.~(\ref{eq:Cb_tilde}) and $\alpha$ was defined in Eq.~(\ref{eq:alpha}). Typically $d>a$, meaning that $\frac{e^2n^2}{\varepsilon}d\gg C_b, C_s$, which implies $\alpha \ll 1$. Then, the onset of the hexatic phase depends only slightly on the gate distance $d$. Thus, the often-made assumption that the Wigner crystal is incompressible ($\tilde{C}_b \rightarrow\infty$) is to a good approximation correct when studying the solid-to-hexatic transition, even in the presence of gates. 

Note that the KTHNY theory is known to describe well the melting of Wigner crystals for densities much lower than the critical density \cite{Grimes_1980}, but it ceases to be applicable for higher densities due to quantum fluctuations \cite{Hwang2001,Clark2009}. In the case of the rhombohedral graphene, taking into account the softening of the lattice at higher temperatures as derived in \cite{Morf1979}, we predict a melting temperature $T_c\sim7$~K, an order of magnitude larger than the observed one $T_c\sim400$~mK \cite{han2026evidencemetallicwignercrystal}. This suggests that the Wigner crystal observed in \cite{han2026evidencemetallicwignercrystal} lies deep in the quantum regime. Nevertheless, a hexatic phase is still expected in some range of temperatures \cite{Clark2009,Bruun2014}.

Let us consider in the following a zero-temperature version of the hexatic Wigner crystal.
The action for the hexatic can be written down using a set of complex scalar order parameters $\Psi_i(t,\vct r)$ transforming as $\Psi_i \rightarrow e^{i\lambda_i^B}\Psi_i$ under the gauge transformation $B_{i\mu}\rightarrow B_{i\mu}+\partial_\mu \lambda_i^B$, see Eq.~(\ref{eq:gauge}). The proliferation of dislocations is then modelled as a spontaneous breaking of this gauge symmetry, where $\Psi_i$ develops a nonzero expectation value $|\Psi|_0$. After integrating out the amplitude oscillations, the remaining terms depend on the phases $\phi_i$ of $\Psi_i$. Denoting
\begin{equation}
    D_{it} \equiv \epsilon_{ij}\left(\partial_t \phi_j - B_{jt}\right),\qquad D_{ij}\equiv \partial_i \phi_j - B_{ij},
\end{equation}
the hexatic Wigner crystal action can be written as
\begin{equation}
\begin{split}
    S_{\mathrm{hex}} =   
    &\frac12 
    \int  d t  d^2k~ \frac{2\varepsilon\, |\vct k|}{\tanh(|\vct k|d)}   A(\vct k)A(- \vct k)\\
    +&\int dt d^2x \left[-\frac{1}{2n_0m}\pi_i^2+\frac{1}{2}\tilde{C}_{ijkl}\tau_{ij}\tau_{kl}\right.\\
    &\left.+\frac{|\psi|_0^2}{2m_1}D_{ii}^2+\frac{|\psi|_0^2}{2m_2}\left(D_{\langle xx\rangle}^2+D_{\langle xy\rangle}^2\right)\right.\\
    &\left. -\frac{1}{2g}D_{it}^2 +C_{\mu}J^{\mathrm{dscl}}_{\mu}\right],
    \label{eq:action_hex1}
\end{split}
\end{equation}
where $m_1$, $m_2$ are parameters with the dimension of mass and $g$ quantifies the strength of local dislocation-dislocation interaction. Furthermore, the gauge fields $A$, $B_{\langle xy\rangle}$ and $B_{\langle xx\rangle}$ are massive, and as such they can be integrated out; then, the low-energy action reads
\begin{equation}
\begin{split}
    \tilde{S}_{\mathrm{hex}} =
    &\int dt d^2x \left[-\frac{1}{2n_0m}\pi_i^2+\frac{1}{2 C_s}\left(\tau_{\langle xy\rangle}^2+\tau_{\langle xx\rangle}^2\right)\right.\\
    &\left.-\frac{1}{4\tilde{C}_b(k)}\tau_{ii}^2+\frac{|\psi|_0^2}{2m_1}D_{ii}^2-\frac{1}{2g}D_{it}^2 +C_{\mu}J^{\mathrm{dscl}}_{\mu}\right],
    \label{eq:action_hex2}
\end{split}
\end{equation}
where the electrostatic potential field $A$ is now absent from the definitions of $\pi_i$ and $\tau_{ij}$ in Eq.~(\ref{eq:field_strengths}), while $B_{\langle ij\rangle}$ are replaced by $\partial_{\langle i}\phi_{j\rangle}$. Thus, all the fields in the Hamiltonian are now expressed through combinations of $C_\mu$, $\phi_i$, and $B_{[xy]}=\frac{1}{2}\left(B_{xy}-B_{yx}\right)$.

In the absence of disclinations, two low-energy modes can be found.
Firstly, there is the longitudinal plasmon mode that couples $\pi_x$, $\tau_{\langle xx\rangle }$, $\tau_{ii}$ and $D_{xt}$, with long-wavelength dispersion 
\begin{equation}
    \omega_{plasm} = \sqrt{\frac{\tilde{C}_b(k)}{2n_0 m}}k
    \label{eq:plasmon2}
\end{equation}
similar to Eq.~(\ref{eq:plasmon}).
The second mode is transverse and couples $\pi_y$, $\tau_{\langle xy \rangle}$, $D_{yt}$ and $D_{ii}$. Its long-wavelength dispersion is linear,
    \begin{equation}
        \omega_{trans} = \sqrt{\frac{g |\psi|_0^2}{m_1}} k.
    \end{equation}
This calculation confirms that the longitudinal plasmon mode is of a purely electronic origin and is only slightly affected by the loss of shear resistance. On the contrary, the transverse mode is charge-neutral and thus it is not affected by the electronic character of the hexatic. This second mode can be understood as the Goldstone mode arising from spontaneous breaking of rotational symmetry, and it has been known before \cite{Zaanen_2004}. It is this rotational Goldstone mode that carries the long-range interactions between disclinations: to the leading order we find that for a point disclination with Frank vector $F$, $\rho^{\mrm{dscl}}(\vct r)=F\delta^{(2)}(\vct r)$,
\begin{equation}
    C_t(\vct r) = -\frac{2g F}{\pi}\ln r.
\end{equation}
Thus, the physics of disclinations, and consequently the hexatic-liquid phase transition, is seemingly unaffected by the electronic nature of the hexatic.

\section{Vacancy proliferation: metallic Wigner crystal}
\label{sec:vacancy}

In Ref.~\cite{han2026evidencemetallicwignercrystal} experimental evidence was presented that under the right conditions, the ground state of the electrons in rhombohedral multilayer graphene can be a Wigner crystal with a finite density of vacancies (holes) $\rho_v$ up to around $0.15 n_0$. According to Table~\ref{tab:interactions}, the interaction energy between such quasiparticles decays exponentially or as $r^{-3}$, depending on whether gate screening is present or not. The $r^{-3}$ scaling is also typical for the Thomas-Fermi screened interaction in metals. The difference, however, is that isolated electrons/holes in a semiconductor polarize their environment, renormalizing the electric permittivity $\epsilon_0\rightarrow \epsilon_r\epsilon_0$ but keeping the Coulombic $r^{-1}$ character of the interaction, which enables a breakdown of the Fermi liquid and formation of an electron crystal. In contrast, even dilute vacancies or interstitials in Wigner crystals will interact via the $r^{-3}$ potential because the lattice of electrons is capable of screening the charge completely. Thus, the low-density defects are generically expected to form a Fermi liquid \cite{Lu2012} rather than a secondary crystal. 

With that in mind, let us explore the consequences of the proliferation of charged defects on the properties of the crystal. In App.~\ref{app:FL-WC} we use the semiclassical Boltzmann equation approach based on the wavepacket construction~\cite{Sundaram1999} to derive the hydrodynamic equations for the ideal Fermi liquid: the continuity equation for charge density and the Euler equation for momentum density. Both hydrodynamic equations can simultaneously be encoded in the low-energy action after adding a new dynamical scalar field $\phi$:
\begin{equation}
\begin{split}
    S_{mWC} =  &\frac12 
    \int  d t  d^2k~ \frac{2\varepsilon\, |\vct k|}{\tanh(|\vct k|d)}   A(\vct k)A(- \vct k)\\  
    + &\frac12 \int  dt d^2x
    \left[n_0m(\partial_t u_i)^2-C_{ijkl}\partial_i u_{j}\partial_ku_{l} \right.\\ &\left.- 2en_{\mrm{eff}}u_iE_i-e^2B\left(\partial_i\phi\right)^2\right. \\
    &\left.  +F \left(eA-e\partial_t\phi+D\partial_iu_i\right)^2-FD^2(\partial_iu_i)^2\right].
    \label{eq:actionWCFL}
\end{split}
\end{equation}
Because vacancies are dilute, we assume that their presence does not renormalize the effective electron mass $m$ and the elastic moduli of the crystal $C_s$, $C_b$. Assuming a parabolic dispersion relation for the holes with effective mass $m_h$ and Fermi energy $\epsilon_F$, we find for the remaining coefficients
\begin{equation}
\begin{split}
    & n_{\mrm{eff}}  = n_0-\rho_v\left(1+\frac{m}{m_h}\right), \qquad  B  = \frac{\rho_v}{m_h},\\
    & F = \frac{\rho_v}{\epsilon_F}, \qquad D  = -\left(1+\frac{m}{m_h}\right)\epsilon_F,
    \label{eq:WCFLcoeffs}
\end{split}
\end{equation}
where $e\rho_v=em_h\epsilon_F/(2\pi\hbar^2)$ is the equilibrium density of $\rho^{\mrm{vac}}$. According to \cite{dong2026crystalscaughtdopingmetallic}, the holes are very light: $m_h\approx 0.05m$. 

The factors $(1+m/m_h)$ that appear in Eq.~(\ref{eq:WCFLcoeffs}) deserve a longer discussion. As discussed in~\cite{Sundaram1999,Holstein1959,Khan1984}, they originate in adiabatic effects induced by the application of strain: the movement of the lattice changes the local band dispersion, modifying the local charge density and giving rise to a charge current. The strength of this effect is proportional to $(1-m/m^*)$, such that for example in the free electron case, $m^*=m$, the motion of conduction electrons and the lattice are decoupled. On the other hand,~\cite{Holstein1959,Khan1984} showed that electron-phonon collisions give rise to an intraband current that equilibrates the velocities of the charged carriers and the lattice, effectively canceling the $-m/m^*$ term. However, at low temperatures $\sim 100$~mK electron-phonon scattering is strongly suppressed and we expect the electron-phonon mean free path $l_{e-ph}$ to be much longer than either the typical device size $\sim1~\mu$m or the characteristic length scale associated with the pinning of the Wigner crystal, which in bilayer graphene has been experimentally estimated at $\sim2~\mu$m \cite{Seiler2025}. Consequently, we shall keep the factors $(1-m/m^*)=(1+m/m_h)$ in Eq.~(\ref{eq:WCFLcoeffs}).

We now integrate out the $A$ field in Eq.~(\ref{eq:actionWCFL}). For the parameters in \cite{han2026evidencemetallicwignercrystal,dong2026crystalscaughtdopingmetallic} we have $\frac{e^2\rho_v}{\epsilon_F}=\frac{e^2m_h}{2\pi\hbar^2}\gg \frac{2\varepsilon k}{\tanh(kd)}$ up to $kd\lessapprox 5$. This means that  screening is primarily due to the vacancies rather than the gates. With this approximation, we get
\begin{equation}
\begin{split}
    \tilde{S}_{mWC} =&   
    \frac12 \int  d\omega d^2k
    \left[-n_0m\omega^2 u_i^2+C_{ijkl}k_ik_k u_{j}u_{l} \right.\\ &\left.+ \frac{2\pi\hbar^2}{m_h}n_0^2(\vct k\cdot\vct u)^2-2en_{\mrm{eff}}(\vct k\cdot\vct u)\omega\phi \right.\\ &\left. -\frac{2\varepsilon k}{\tanh(kd)}\omega^2\phi^2+\frac{e\rho_v}{m_h}k^2\phi^2\right].
    \label{eq:actionWCFL2}
\end{split}
\end{equation}

First, let us look at the static properties of the metallic Wigner crystal by setting $\omega=0$. We see that $\tilde{S}_{mWC}$ describes a charge-neutral crystal with the bulk elastic modulus renormalized as $C_b\rightarrow C_b+n_0^24\pi\hbar^2/m_h$. Using the elastic constants in~\cite{Tozzini_1996}, we find that assuming the classical KTHNY theory to be applicable, the critical temperature for the crystal-to-hexatic transition~(\ref{eq:kbThexatic}) should drop by about $10\%$ compared to the pure Wigner crystal. However, as noticed in the discussion below Eq.~(\ref{eq:kbThexatic}), the metallic Wigner crystal reported in~\cite{han2026evidencemetallicwignercrystal} is presumably deep within the quantum regime, which might explain why the melting temperatures reported for both the pure Wigner crystal and the metallic Wigner crystal are very similar. More studies would be necessary to understand this phenomenon.

\begin{figure}
    \includegraphics[width=0.8\linewidth]{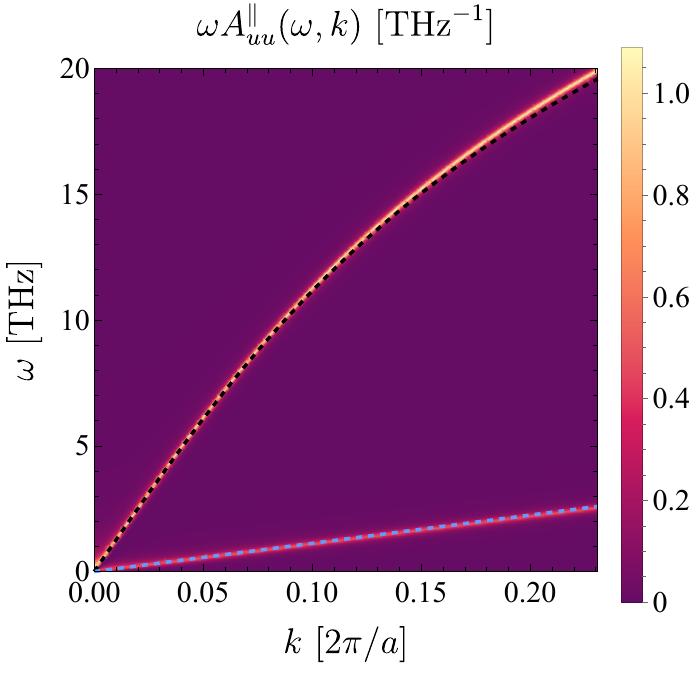}
    \caption{Spectral function of the longitudinal mode $\vct u \parallel\vct k $ at $\rho_v=0.06n_0$. The spectral function is defined as $A^{\parallel}_{uu}(\omega)=-\pi^{-1}\mrm{Im}[G^{\parallel}_{uu}(\omega-i\gamma)]$, where $G^{\parallel}(\omega)$ is the Green's function matrix for the longitudinal $\vct u \parallel\vct k $ sector of the action~(\ref{eq:actionWCFL2}) and $\gamma$ is the imaginary part of the self energy. To illustrate the relative spectral weight of the two modes, we set a constant $\gamma=10^{11}$ Hz. The thin black dashed line follows $\omega_{\mrm{plasm}}(k)$, and the think blue dashed line follows $\omega_{\mrm{ph}}(k)$ in Eq.~(\ref{eq:WCFLmodes}). We stress that the quasiparticle decay rate $\gamma$ is chosen arbitrarily, but the spectral weight of the modes does not depend on $\gamma$. In particular, we do not consider the particle-hole continuum effects.}
    \label{fig:spectral}
\end{figure}

\begin{figure}
    \includegraphics[width=\linewidth]{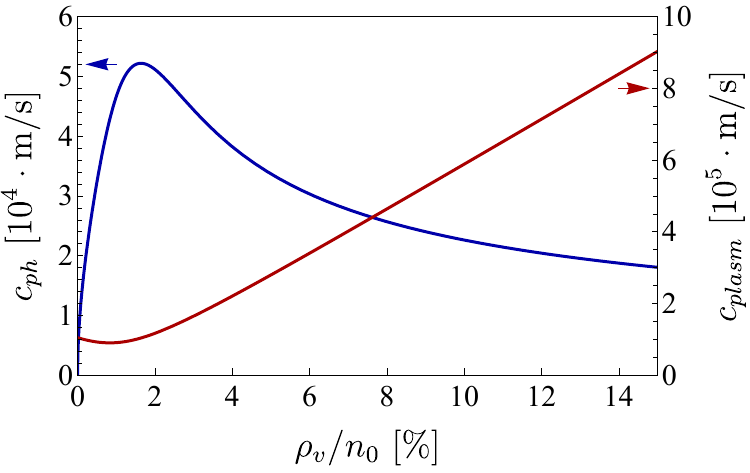}
    \caption{The speed of sound of the two modes in Eq.~(\ref{eq:WCFLmodes}) for $k\ll d^{-1}$ as a function of the relative density of holes, $\nu=\rho_v/n_0$. Note the different scales on the $y$-axes.}
    \label{fig:speeds}
\end{figure}

Next, let us approximate the long-wavelength longitudinal phonon spectrum of the theory. The action in Eq.~(\ref{eq:actionWCFL2}) produces two longitudinal modes with:
\begin{equation}
    \begin{split}
        \omega_{\mrm{plasm}}^2(k) & \approx \frac{e^2}{2\varepsilon}\left(\frac{\rho_v}{m_h}+\frac{n_{\mrm{eff}}^2}{n_0m}\right)\tanh(kd)k, \\
        \omega_{\mrm{ph}}^2(k) & \approx \frac{C_b+C_s+4\pi\hbar^2n_0^2/m_h}{2n_0m+2n_{\mrm{eff}}^2m_h/\rho_v}k^2.
    \end{split}
    \label{eq:WCFLmodes}
\end{equation}
In Fig.~\ref{fig:spectral} we plot the spectral function for $\rho_v=0.06n_0$ with an arbitrary quasiparticle decay rate. We find that for this $\rho_v$, the phonon mode has a lower spectral weight than the plasmon, with the speed of sound $c_{ph}\approx 30$ km/s, same order of magnitude as the acoustic phonon in graphene. Curiously, when $\rho^*_v=n_0/(1+m/m_h)$, then $n_{\mathrm{eff}}=0$, around which point the phonon speed $c_{\mathrm{ph}}$ reaches its maximum, while the plasmon speed $c_{\mathrm{plasm}}$ has a minimum, see Fig.~\ref{fig:speeds}. We propose that thanks to this mechanism, measuring the speed of sound can be used to study properties of the vacancies. By comparing the measured $c_{\mrm{ph}}$ with the curve in Fig.~\ref{fig:speeds}, one can estimate $\rho_v$ for the different experimental parameters. Furthermore, at $\rho_v^*$ we find $c_{\mrm{ph}}^2\approx 2\pi\hbar^2 n_0/(mm_h)$ and thus the density of states of the vacancy fluid $m_h/(2\pi\hbar^2)$ can in principle be extracted from a speed of sound measurement. Alternatively, if the density $\rho_v$ can be measured independently but the effective electron mass is not known, the same set of measurements can be used to determine both $m$ and $m_h$.

\section{Discussion}

We have developed fracton-elasticity duality for two-dimensional crystals interacting electrostatically in the presence of metallic gates. We have found that within the classical regime, the thermal melting can be described by the standard KTHNY theory with the bulk elastic modulus being effectively infinite, as in earlier treatments \cite{Fisher1979,Morf1979,Esfarjani1992}. Self-doping of vacancies or interstitials is predicted to lead to a metallic state, in agreement with \cite{Barabanov1995,Barraza1999,Pankov2008,Kivelson2024,han2026evidencemetallicwignercrystal,dong2026crystalscaughtdopingmetallic,feng2026selfdopedcrystalpreemptedbandinversion}. While we expect that the melting temperature is lower in the metallic Wigner state that in the bare crystal, the similar melting temperatures observed in \cite{han2026evidencemetallicwignercrystal} are also consistent with our predictions close to the quantum melting point.

At the same time, many of the results obtained here are novel. We have for the first time obtained analytical formulas for interactions between pairs of defects other than two dislocations. We have considered the collective modes in a hexatic Wigner crystal. Finally, we have analyzed how the speed of sound of the longitudinal acoustic phonon in the metallic Wigner crystal depends on the density and effective mass of the vacancies or interstitials, suggesting an experimental way to measure them.

The utility of the framework and the results obtained here is not limited to Wigner crystals. Rather, it is applicable to any charged system where the translational symmetry is spontaneously broken. One example of this is a charge density wave, in which case the phase of the condensate plays a role analogous to the displacement in the Wigner crystal. Another example is the recently proposed anomalous Hall crystals; while their complete description requires incorporating quantum geometry, it does not directly affect the collective modes in the long-wavelength limit \cite{Zeng2025,Tomo2025}.

These results have been obtained for a simplified model that neglects particle spin, statistics, and pinning due to impurities. 
Including such effects is possible in the presented field theory, and is expected to further close the gap between theory and experiment.
Another worthwhile direction is to consider the interactions of the phonons of the crystal with excitons, which is known to give rise to additional collective modes \cite{zhang2025wignerpolaronsrevealwigner,wang2025spectroscopywignercrystalpolarons,adlong2025theoryexcitonpolarons2d,nyhegn2025excitoninteractingphononselectronic}. Finally, the fact that the speed of sound of the metallic Wigner crystal is comparable to the ionic speed of sound in graphene suggests that a more complete description of this phase might require coupling all three types of charged carriers, i.~e.\ ionic cores, conduction band electrons and vacancies, on equal footing.

\section{Acknowledgements}

We acknowledge financial support by the European Research Council (ERC) under grant QuantumCUSP (Grant Agreement No.\ 101077020). P.M. acknowledges support by the Harry Bloomfield International Scholarship and a postdoctoral fellowship from the Azrieli Foundation.

\appendix
\setcounter{figure}{0}
\renewcommand{\thefigure}{A\arabic{figure}}

\section{Effective action for the mixed Fermi liquid-Wigner crystal system}
\label{app:FL-WC}

To describe the dynamics of a Fermi liquid of holes in a dynamical background provided by the Wigner crystal, we resort to the semiclassical wavepacket treatment. Following Ref.~\cite{Sundaram1999}, the equations of motion for the wavepacket center of mass $\vct r$ and center of momentum $\vct k$ are
\begin{equation}
\begin{split}
    \dot{r}_i & = \frac{\partial \epsilon(\vct k)}{\partial k_i}+\partial_k u_j\frac{\partial D_{jk}(\vct r,\vct k)}{\partial k_i}+\mathcal{M}_{ij}(\vct k)\left(\partial_tu_j+\dot{r}_k \partial_k u_j\right), \\
    \dot{k}_i & = -D_{jk}(\vct r,\vct k)\partial_i\partial_k u_j+\partial_i u_j \mathcal{M}_{jk}(\vct k) \dot{k}_k -e E_i,
\end{split}
\end{equation}
where $u_i$ is the crystal displacement field, $\epsilon(\vct k)$ is the dispersion relation for zero displacement field $u_i=0$, and
\begin{equation}
\begin{split}
    D_{ij}(\vct r,\vct k) & = \frac{\partial \Delta\epsilon(\vct r,\vct k)}{\partial (\partial_j u_i)}, \\
    \mathcal{M}_{ij}(\vct k) & = \delta_{ij}-\frac{m\partial^2\epsilon(\vct k)}{\partial k_i \partial k_j},
\end{split}
\end{equation}
where $\Delta \epsilon(\vct r,\vct k)$ is the correction to energy due to crystal deformation (discussed in more detail in~\cite{Sundaram1999}), and $m$ is the effective mass of the Wigner crystal electron. Within the linear order in crystal deformation, $\mathcal{O}(\partial_i u_j)$, we find
\begin{equation}
\begin{split}
    \partial_i \dot{r}_i + \partial_{k_i} \dot{k}_i = & \mathcal{M}_{ij}(\vct k) \left(\partial_t\partial_iu_j+\dot{r}_k \partial_k \partial_iu_j\right)\\
    &+\frac{\partial \mathcal{M}_{jk}(\vct k)}{\partial k_i}\dot{k}_k \partial_i u_j +\mathcal{O}((\partial_i u_j)^2) \\
    = & \frac{d}{dt}\left(\mathcal{M}_{ij}(\vct k) \partial_iu_j\right) +\mathcal{O}((\partial_i u_j)^2),
\end{split}
\end{equation}
where we used $\partial \mathcal{M}_{jk}/\partial k_i=-m\partial^3\epsilon/(\partial k_i\partial k_j\partial k_k)=\partial \mathcal{M}_{ij}/\partial k_k$. This shows that the invariant volume element $\mathcal{V}(\vct k)d^2rd^2k$ is
\begin{equation}
    \mathcal{V}(\vct k)d^2rd^2k = \left(1-\mathcal{M}_{ij}(\vct k) \partial_iu_j\right)d^2rd^2k,
\end{equation}
because $d\left(\mathcal{V}(\vct k)d^2rd^2k\right)/dt=0$. Now we will assume that the charge carriers are holes and express the distribution function of the electrons $f(\vct r,\vct k)$ through the distribution function of the holes $f_v(\vct r,\vct k)$ as $f=(1-f_v)$ and denote the dispersion of the holes as $\epsilon_v(\vct k)$. In the linear response regime, we expand the distribution function as $f_v=f_0+(\partial_{\epsilon}f_0)\delta f$, where $f_0(\vct k)$ is the Fermi-Dirac distribution and $\delta f(\vct r, \vct k)$ is the correction to it linear in the electric field $\vct E$ and strain $\partial_iu_j$. The electric charge and current densities are given, to the first order, by
\begin{align}
    \begin{split}
    \rho^{\mrm{vac}}&=e\int \frac{d^2k}{(2\pi)^2}\mathcal{V}(\vct k)f_v \\ 
    &= e\rho_v-e\rho_v\left(1+\left\langle\frac{m}{m^*} \right\rangle\right)\partial_i u_i+e\delta \rho(\vct r), \label{eqapp:charge_orig}
    \end{split}\\
    \begin{split}
    J^{\mrm{vac}}_i& =e\int \frac{d^2k}{(2\pi)^2}\mathcal{V}(\vct k)\dot{r}_if_v \\ & =e\rho_v\left(1+\left\langle\frac{m}{m^*} \right\rangle\right)\partial_tu_i+ e\delta J_i(\vct r),\label{eqapp:current_orig}
    \end{split}
\end{align}
where
\begin{equation}
\begin{split}
    \rho_v & \equiv \int \frac{d^2k}{(2\pi)^2}f_0(\vct k),\\
    \left\langle\frac{1}{m^*} \right\rangle & \equiv \rho_v^{-1}\int \frac{d^2k}{(2\pi)^2}\frac{\partial^2\epsilon_v(\vct k)}{\partial k_i \partial k_i} f_0(\vct k),\\
    \delta\rho(\vct r) & \equiv \int \frac{d^2k}{(2\pi)^2}\frac{\partial f_0(\vct k)}{\partial \epsilon_v}\delta f(\vct r, \vct k),\\
    \delta J_i(\vct r) & \equiv \int \frac{d^2k}{(2\pi)^2}\frac{\partial f_0}{\partial k_i}\delta f(\vct r,\vct k).
\end{split}
\end{equation}
In the derivation, we have used the hexagonal symmetry to constrain the allowed tensor structures: $\left\langle\frac{1}{m^*} \right\rangle_{ij}\equiv \left\langle\frac{1}{m^*} \right\rangle \delta_{ij}$ and $\int_{\vct k}\mathcal{M}_{ij}\frac{\partial \epsilon}{\partial k_j}f_0=0$ and $\int_{\vct k}\frac{\partial D_{ij}}{\partial k_j}f_0=0$.

In Eq.~(\ref{eqapp:current_orig}), there appears the factor $\left(1+\left\langle\frac{m}{m^*} \right\rangle\right)$ that modifies the charge density and current. It originates from the adiabatic change of the band dispersion as the crystal is strained and can be interpreted as an interband effect. On the other hand, as outlined in \cite{Holstein1959,Khan1984}, this factor gets corrected in the presence of electron-phonon scattering. Let us denote the typical scattering rate as $\Gamma$ and mean free path as $l$.  If the frequency $\omega$ and wavevector $\vct q$ of the perturbation are small enough such that $\omega\ll\Gamma$, $q\ll2\pi/l$, the Fermi-Dirac distribution receives a correction $f_0(\vct k)\rightarrow f_0(\vct k-m\partial_t \vct u_i )$. This correction precisely cancels the $-e\rho_v\left\langle\frac{m}{m^*} \right\rangle\partial_t\vct u$ contribution to Eq.~(\ref{eqapp:current_orig}). The continuity equation then dictates that the electron density must also adjust to cancel the $e\rho_v\left\langle\frac{m}{m^*} \right\rangle\nabla\cdot\vct u$ contribution to Eq.~(\ref{eqapp:current_orig}). Thus, when discussing the static limit, one should take
\begin{align}
    \rho^{\mrm{vac,st}}&= e\rho_v-e\rho_v\partial_i u_i+e\delta \rho(\vct r), \label{eqapp:charge}\\
    J^{\mrm{vac,st}}_i& =e\rho_v\partial_tu_i+ e\delta J_i(\vct r),\label{eqapp:current}
\end{align}
meaning the vacancies are fully dragged along by the moving Wigner crystal. 

To find the equations of motion for the Fermi surface quantities $\delta\rho$ and $\delta J_i$, we use the Boltzmann equation,
\begin{equation}
    \partial_t f + \dot{r}_i\partial_i f+\dot{k}_i\partial_{k_i}f = C[f], \label{eqapp:boltzmann}
\end{equation}
where $C[f]$ is the collision integral. In the absence of momentum dissipation, we have the conservation laws
\begin{equation}
    \int \frac{d^2k}{(2\pi)^2}\mathcal{V}(\vct k)C[f]=0,\qquad \int \frac{d^2k}{(2\pi)^2}\mathcal{V}(\vct k)C[f]k_i=0.
\end{equation}
For simplicity we will also assume that the non-conserved quantities relax instantaneously, meaning that we treat the Fermi liquid as an ideal fluid. Then, the continuity equations can be found by multiplying Eq.~(\ref{eqapp:boltzmann}) by factors $1$ and $\partial \epsilon_v(\vct k)/\partial k_i$, respectively, and integrating over $\vct k$. 
We further simplify by restricting to a parabolic dispersion relation
\begin{equation}
    \epsilon_v(\vct k) = \frac{k^2}{2m_h},
\end{equation}
and we get to the first order in perturbation:
\begin{align}
    \partial_t\delta\rho + \partial_i\delta J_i & = 0,\label{eqapp:cont1}\\
    \partial_t\delta J_i+\frac{1}{2}v_F^2 \partial_i\delta\rho-\frac{e\rho_v}{m_h} E_i+\frac{D\rho_v}{m_h}\partial_i\partial_ju_j & = 0.\label{eqapp:cont2}
\end{align}
Here, $v_F$ is the Fermi velocity and $D$ is the derivative of the total energy of the Fermi liquid with respect to $\partial_iu_i$, which using Eq.~(\ref{eqapp:charge_orig}) can be written as
\begin{equation}
    D = -\left(1+\frac{m}{m_h}\right)\epsilon_F.
\end{equation}
Now we notice that the first-order continuity equations in Eq.~(\ref{eqapp:cont1})-(\ref{eqapp:cont2}) follow from the following action:
\begin{equation}
\begin{split}
    S_{\mrm{vac}}[\phi] = & \frac{\rho_v}{2m_h}\int  dt d^2x\left[\frac{2}{v_F^2}\left(eA-e\partial_t\phi+D\partial_iu_i\right)^2\right. \\
    &\left.-e^2\left(\partial_i\phi\right)^2-\frac{2}{v_F^2}D^2(\partial_iu_i)^2\right].
    \label{eqapp:Lfree}
\end{split}
\end{equation}
Indeed, the equations of motion $\delta S_{\mrm{vac}}[\phi]/\delta\phi=0 $ are equivalent to Eq.~(\ref{eqapp:cont1}) upon identification
\begin{equation}
\begin{split}
    \delta\rho & = -\frac{2\rho_v}{m_hv_F^2} \left(eA-e\partial_t\phi+D\partial_iu_i\right),\\
    \delta J_i & = -\frac{e\rho_v}{m_h}\partial_i\phi,
\end{split}
\end{equation}
while Eq.~(\ref{eqapp:cont2}) is satisfied as an identity. The last term in Eq.~(\ref{eqapp:Lfree}) was added to make the action nonnegative definite, but it can also be interpreted as the energy density of the Fermi liquid under strain: note that $\frac{2\rho_v}{m_hv_F^2}=\frac{\rho_v}{\epsilon_F}$ is the density of states. In the end we conclude that, at first order in gradient expansion, the effect of Fermi liquid can be captured by adding~(\ref{eqapp:Lfree}) to the action~(\ref{eq:action2D}) and correcting the effective charge density of the Wigner crystal $en_0\partial_iu_i$ by the terms found in Eq.~(\ref{eqapp:charge_orig}) or~(\ref{eqapp:charge}). 

\bibliography{quadrupole}

\end{document}